\documentclass[12pt]{article} 

\usepackage{PRIMEarxiv}
\usepackage[utf8]{inputenc} 
\usepackage[T1]{fontenc}    

\usepackage{hyperref}       
\usepackage{url}            
\usepackage{booktabs}       
\usepackage{amsfonts}       
\usepackage{nicefrac}       
\usepackage{microtype}      
\usepackage{graphicx}       
\usepackage{lipsum}         
\usepackage{times}          
\usepackage{setspace}       
\usepackage{amsmath}
\usepackage{relsize}   
\usepackage{pdfpages}

\usepackage{anyfontsize}       
\usepackage[font=small,labelfont=bf]{caption} 

\renewcommand\normalsize{\fontsize{11}{15}\selectfont}

\title{\textbf{Ultra-miniaturized Bloch mode metasplitters for one-dimensional grating waveguides}}

\author{
  Ahmet Oguz Sakin$^{1}$, Hamza Kurt$^2$, Mehmet Unlu$^{1*}$ \\
  $^1$Department of Electrical and Electronics Engineering, TOBB University of Economics and Technology, Ankara, Turkey \\
  $^2$School of Electrical Engineering, Korea Advanced Institute of Science and Technology (KAIST), Daejeon, Republic of Korea \\
  \textbf{*Corresponding author:} \texttt{munlu@etu.edu.tr} \\
}

\makeatletter
\renewcommand{\maketitle}{
  \begin{center}
    {\LARGE \textbf{\@title} \par}
    \vspace{1em}
    {\large \@author}
  \end{center}
}
\makeatother
\begin{document}
\maketitle

\begin{abstract}
We present, for the first time, power splitters with multiple channel configurations in one-dimensional grating waveguides (1DGWs) that maintain crystal lattice-sensitive Bloch mode profiles without perturbation across all output channels, all within an ultra-miniaturized footprint of just 2.1 $\times$ 2.2 $\mu\text{m}^2$. This novel capability reduces the need for transition regions, simplifies multi-channel configurations of 1DGWs, and maximizes the effective use of chip area. The pixelated metamaterial approach, integrated with a time-domain heuristic algorithm, is utilized to concurrently achieve broadband operation, optimized dispersion control, and minimal loss. We experimentally demonstrate that our 1x2 and 1x3 metasplitters achieve average minimum losses per channel of $3.80 \, \text{dB}$ and $5.36 \, \text{dB}$, respectively, which are just $0.80 \, \text{dB}$ and $0.59 \, \text{dB}$ above ideal splitting. The measurements for both designs demonstrate a 1 dB bandwidth of 15 nm, with excellent uniformity across all output channels. These versatile metasplitter designs can serve as fundamental building blocks for ultrahigh-bandwidth, densely integrated photonic circuits and in scenarios where slow light is essential.
\end{abstract}


\section{Introduction} 
1DGWs are meticulously engineered artificial structures designed to control light propagation by manipulating lattice-sensitive Bloch modes. These waveguides have emerged as promising components in cutting-edge applications, including microwave photonics \cite{burla2013integrated}, signal processing \cite{zhang2018fully}, quantum photonics \cite{xie2020chip}, optical modulators \cite{han2023slow}. In these applications, multi-channel configurations are predominantly utilized to optimize system performance, increase capacity, and enhance scalability. To achieve these goals, power splitting is a key operation in 1DGWs, which is conventionally achieved using multimode interferometers (MMIs) \cite{yi2023silicon}, directional couplers \cite{ye2020ultra}, Y-branches \cite{ozcan2023short}, or photonic crystals \cite{he2020topologically}. However, integrating 1DGWs with traditional power splitters presents challenges due to disparities in dispersion properties and mode profiles. Consequently, channels are initially divided using conventional power dividers before employing 1DGWs, leading to significant chip area consumption and increased system complexity.

\begin{figure*}[!t]
\centering
\includegraphics[height=2.3in,width=6.5in]{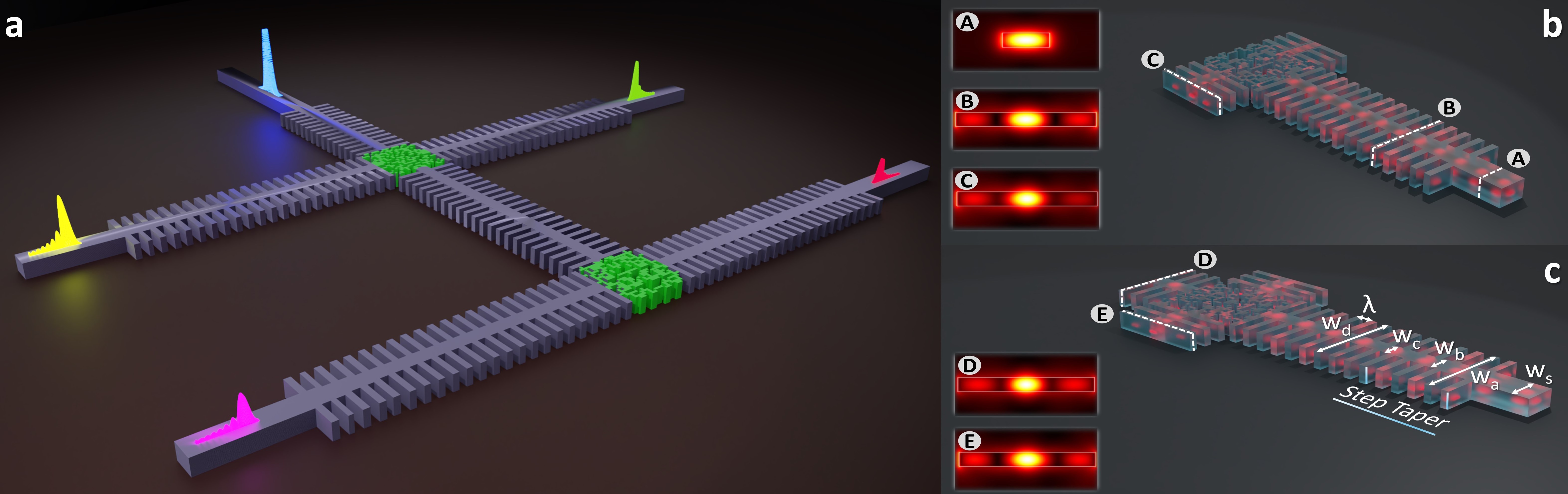}
\caption{(a) Schematics of the 1x2 and 1x3 metasplitters designed for Bloch mode splitting in multi-channel 1DGW-based systems (Different colored outputs highlight the possibility of pulse customization between channels with varying periods). (b) Close-up of the 1x2 metasplitter with 1DGW connections, highlighting Bloch mode splitting capabilities and corresponding mode profiles (A–C). (c) Dimensions of 1DGWs and close-up of the 1x3 metasplitter with corresponding mode profiles (D–E).}
\label{fig:false-color}
\end{figure*}

These challenges are particularly evident in true-time delay line configurations, which demand substantial chip space. Moreover, design complexity in these systems further increases when conventional power splitters are used to expand channels or accommodate complex 2D configurations, especially in antenna array systems \cite{wang2018continuously,chung2018chip, jung2007optical}. A specific problem arises since power splitting does not occur as light propagates through the 1DGW structure, it is necessary to use double-step taper structures—one on the input side and one on the output side—between the strip waveguide and 1DGW structures in each channel. This design requirement presents a significant challenge, especially in scenarios with high group index-based 1DGWs, as achieving perfect coupling ratios in the transitions is difficult \cite{zhao2017efficient}. By implementing power splitting in 1DGWs, the number of step taper structures in a system with $L$ channels can be reduced from $2L$ to $L + 1$. This new configuration includes only one taper structure at the input and the remaining transitions at the outputs. However, even with the emergence of this new configuration concept, the conventional design approach remains challenging for power splitting in 1DGWs. This is due to the high group index sensitivity, strongly wavelength-selective nature, and lattice-sensitive Bloch mode profiles, which are difficult to address when relying on the designer's intuition. Therefore, a substantial challenge persists in developing ultra-compact and low-loss Bloch mode splitters for 1DGWs.

In this Letter, we present, to the best of our knowledge, the first Bloch mode power splitters for 1DGWs, demonstrated in 1x2 and 1x3 configurations, which occupies a footprint of only $2.1 \times 2.2 \, \mu\text{m}^2$. The dependence of Bloch modes on the crystal lattice structure sets them apart from standard photonic design libraries. To effectively address the challenges of light manipulation within this framework, inverse design methods have been rigorously explored \cite{vercruysse2021inverse}. Among these, we prefer the pixelated metamaterial approach, which balances design complexity and simplicity \cite{della2014digital}. This method optimizes key interdependent parameters, including broadband performance, uniformity, and dispersive compatibility using heuristic algorithms.

\begin{figure*}[!t]
\centering
\includegraphics[height=2.3in,width=6.5in]{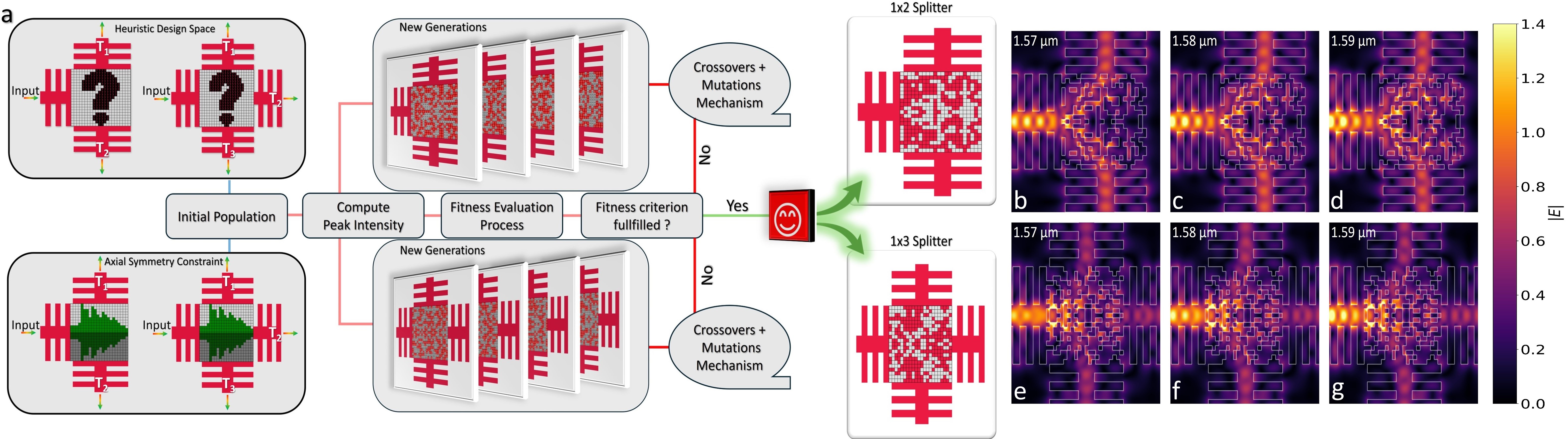}
\caption{(a) Flowchart of the heuristic algorithm used for the inverse design of 1x2 and 1x3 metasplitters. (b–d) E-field distributions of the 1x2 metasplitter at 1.57 µm, 1.58 µm, 1.59 µm. (e–g) E-field distributions of the 1x3 metasplitter at the same wavelengths.}
\label{fig:false-color}
\end{figure*}

Figure 1-(a) depicts a multi-channel 1DGW system that includes both 1x2 and 1x3 pixelated metasplitters. In 1DGWs, light propagates as Bloch mode profiles, consistent with the crystal lattice of the structure, due to the interaction between the light wave and the periodic grating structures. Therefore, the design of ultra-miniaturized metasplitters must ensure the preservation of a consistent Bloch mode profile. Typically, the Bloch field intensity reaches the tips of the grating; thus, designing metasplitters smaller than the grating width could potentially disrupt this Bloch mode profile. In Figure 1-(b), the mode labeled A corresponds to the TE fundamental mode, which, after the transition region, evolves into the mode labeled B, representing the Bloch mode profile, reaching the tip of the grating structures as depicted. As illustrated in Figure 1-(c), the inner waveguide width of the 1DGW is denoted as $w_c$, the width of the corrugated section as $w_d$, and the period as $\Lambda$. Group-velocity mismatch occurs in the butt coupling from the strip waveguide, which has a width $w_s$, to the grating waveguide. To minimize band narrowing and decreases in the coupling ratio, a step taper structure based on the anti-Fresnel reflection method is employed \cite{zhao2017efficient, hosseini2011role}. The step taper design gradually narrows from $w_s$ to $w_c$, as shown in Figure 1-(c), with five distinct widths, denoted as $w_b$. In the toothed segment design, the parameter $w_a$ is uniformly selected to be less than 100 nm, as compared to the corrugated part of the grating waveguide ($w_d$). The 1DGW topology has been optimized to achieve a broadband true-time delay in the L-band, with a group index of 6.2 at 1580 nm (Sect. S1, Supplement I). Therefore, the grating period ($\Lambda$), corrugation width ($w_d$), and inner waveguide width ($w_c$) are set to 380 nm, 2100 nm, and 450 nm, respectively. The device is designed on a silicon-on-insulator (SOI) wafer, consisting of a 220 nm thick silicon device layer and a 2 \textmu m thick buried oxide (BOX) layer. The strip waveguide width ($w_s$) is set to 700 nm to support two TE-like modes and two TM-like modes.

The metasplitter design area is specified as $2.1 \, \mu\text{m} \times 2.2 \, \mu\text{m}$, ensuring that both dimensions exceed the grating width, as previously mentioned. This specification provides the most feasible boundaries for a highly miniaturized design region.  Considering the computational cost of inverse design and specific fabrication constraints, the square pixel size is set to $100 \, \text{nm} \times 100 \, \text{nm}$. This design method indicates that the material properties are binary, consisting of either silicon or silicon dioxide. As shown in Figure 2-(a), the inverse design flowchart for the metasplitters, utilizing the genetic algorithm, is illustrated.

The process starts with an initial random population of designs within a heuristic design space, defined by specific dimensions. To ensure consistent performance across different channels and simultaneously reduce computational cost, an axial symmetry constraint is applied. The most challenging aspect of this work is maintaining low insertion loss while ensuring high uniformity across a broad bandwidth. Therefore, selecting an appropriate Figure of Merit (FoM) is crucial. It is well established that a temporal signal can be designed to encompass all relevant frequencies of interest. Therefore, utilizing temporal signals facilitates the efficient design of metastructures with the desired dispersion characteristics and uniformity across a broadband spectrum. In line with this approach, and considering the spectral characteristics of 1DGWs, an ultra-short signal with a duration of 90 fs at 1580 nm is employed as the input. Evaluating the output signals can determine whether the designed metasplitters have the proper group-velocity profile, uniformity, and broadband characteristics. Problems in these areas often result in temporal broadening and distortion of the ultra-short pulse shape, leading to reduced peak intensity \cite{hickstein2019self} (Sect. S2, Supplement I). Therefore, the suitable FoM of a 1xN power splitter system can be formulated as shown in Equation 1.


\begin{align}
\text{FoM} =\alpha \left[ 1 - 2 \left( \sum_{i=1}^{N_s} S_p^i(t) \right) - S_p^{N_s + 1}(t) \cdot \xi \right] 
+ \beta \sum_{i=1}^{N_s + \xi} \sum_{j=i+1}^{N_s + \xi} \left| S_p^i(t) - S_p^j(t) \right| 
\label{eq:FoM_Speak}
\end{align}

\noindent Here, $S_{\text{p}}^i(t)$ and $S_{\text{p}}^j(t)$ represent the peak values of the time-domain Poynting vector for channels. $\alpha$ and $\beta$ are weighting factors. $N_s$ represents the number of symmetric channels, where $N_s = \frac{N}{2}$ for even $N$ and $N_s = \frac{N-1}{2}$ for odd $N$. Additionally, $\xi = 1$ if $N$ is odd and $\xi = 0$ if $N$ is even. The first term of the FoM ensures overall efficiency, while the second term promotes uniformity among channels. Due to symmetry, only half of the channels are considered for the sums and uniformity. As a result, the FoM for the 1x2 splitter is defined as $\text{FoM} = 1 - 2(S_{\text{p}}^1(t))$, while for the 1x3 splitter it is defined as $\text{FoM} = \alpha[1 - 2S_{\text{p}}^1(t) - S_{\text{p}}^2(t)] + \beta|S_{\text{p}}^1(t) - S_{\text{p}}^2(t)|$.  Each design is evaluated using a fitness check. If it doesn't meet the criteria, new designs are generated through crossovers and mutations.

To assess the designs, we use 3D Finite-Difference Time-Domain (FDTD) simulations, as 2D simulations, though computationally efficient, fall short of accurately capturing losses and mode profiles, especially for the Bloch mode in 1DGWs \cite{vercruysse2019dispersion}. The simulated E-field results show effective light division into two and three channels for both metasplitters, as illustrated in Figure 2-(b) to (g). In Figure 2-(b) to (d), the 1x2 metasplitter separates high and low field intensities of the Bloch mode without mode conversion, as supported by the simulated E-field profile shown in Figure 1-(b), labeled C. For the 1x3 metasplitter, the Bloch mode distributes power into three channels, as shown in Figure 2-(e) to (g), with the Bloch mode profile preserved in the simulated E-field models D and E in Figure 1-(c). 

The metasplitters are fabricated using electron beam lithography, as part of the NanoSOI multi-project wafer process by Applied Nanotools (ANT) \cite{chrostowski2019silicon}. The scanning electron microscope (SEM) images of the meta devices are shown in Figure 3-(a) and (b). To characterize the metasplitters, the input signal is coupled into a grating coupler using a polarization-maintaining fiber (PMF), while the output signal for each channel is guided through a PMF with a grating coupler and subsequently measured with Agilent 81635A optical power sensors, which operate within a bandwidth of 1480-1580 nm. To eliminate the influence of grating couplers on the characterization results, a grating coupler-straight waveguide-grating coupler structure is employed.

Figure 3-(c) presents the simulation result of the 1DGW, which shows a 1 dB bandwidth of 39 nm and a loss of 0.74 dB at 1580 nm. As shown in Figure 3-(d) and (e), the simulated transmission spectra are presented for both the conventional strip waveguide (WG) and the metasplitter-based configurations used for power splitting. In Figure 3-(d), the 1x2 splitting performance is significantly improved with the metasplitter design, reducing the loss from 11.74 dB to 4.83 dB at 1580 nm, while also increasing the bandwidth. Similarly, Figure 3-(e) illustrates a more uniform 1x3 splitting operation across a broader wavelength range, with transmission ratios at 1580 nm—initially -13.71 dB, -3.73 dB, and -13.71 dB for the channels in the conventional strip waveguide configuration—improved to -6.48 dB, -6.09 dB, and -6.48 dB using the metasplitter design.

\begin{figure*}[!t]
\centering
\includegraphics[height=4.5in,width=6.5in]{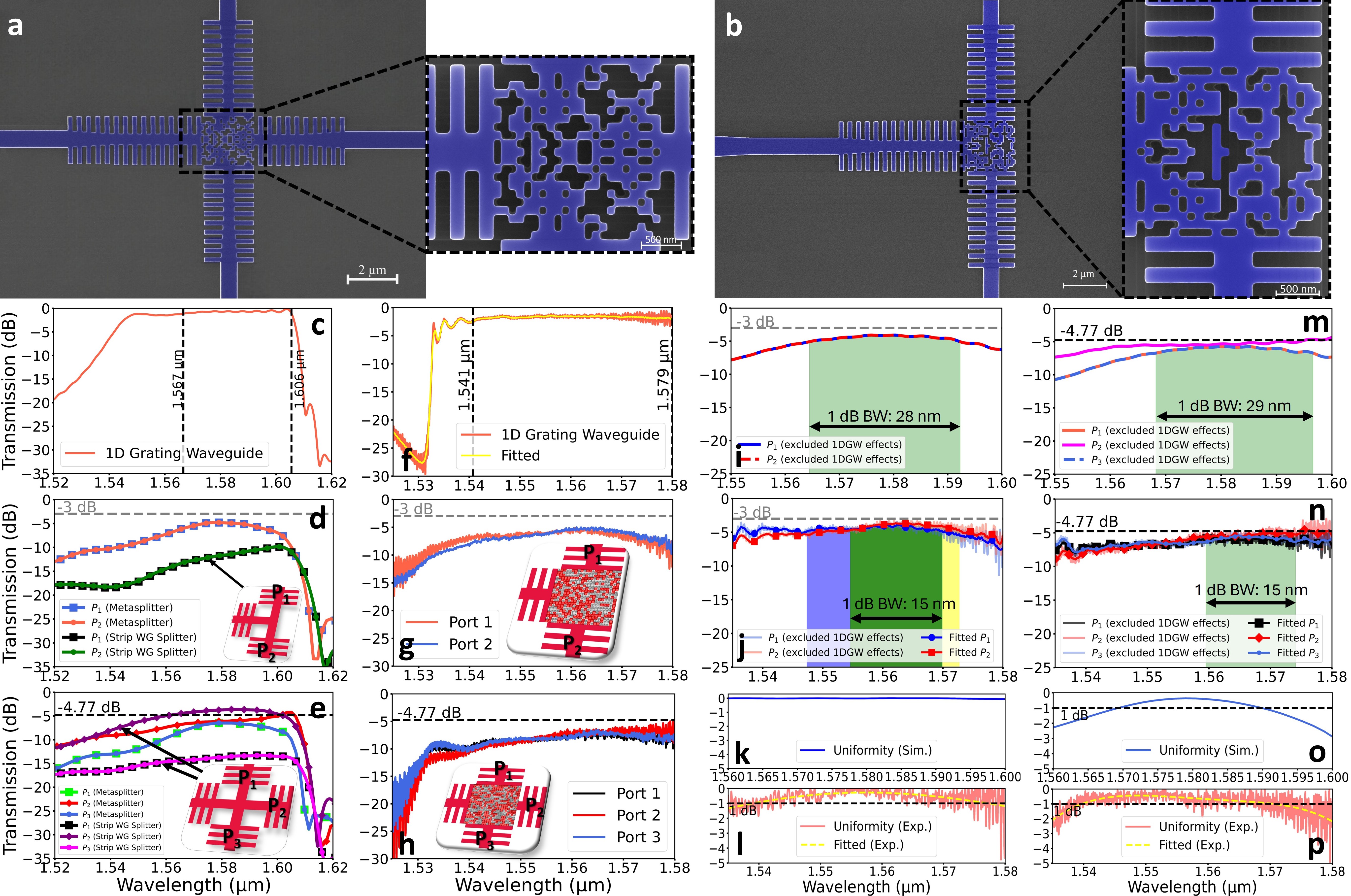}
\caption{(a-b) SEM images of the 1x2 and 1x3 metasplitters. (c-e) Simulated transmission spectra for the 1DGW, strip waveguide, and metastructure-based splitters for 1x2 and 1x3 configurations. (f-h) Measured transmission spectra for the 1DGW, 1x2, and 1x3 metasplitters. (i-j) Simulated and measured intrinsic transmission results for the 1x2 metasplitter. (k-l) Simulated and measured uniformity for the 1x2 metasplitter. (m-n) Simulated and measured intrinsic transmission results for the 1x3 metasplitter. (o-p) Simulated and measured uniformity for the 1x3 metasplitter. In (o, p, l), the dashed line indicates the 1 dB level.}
\label{fig:false-color}
\end{figure*}

The measurement results for the 1DGW, 1x2, and 1x3 metasplitters are presented in Figure 3-(f) to (h), respectively. As shown in Figure 3-(f), the 1 dB bandwidth of the 1DGW is measured to be 38 nm, remaining nearly consistent with the simulation, with only a 1 nm reduction. Additionally, an average band shift of 26 nm is observed between the experimental and simulation results. Consequently, the simulated loss of 0.74 dB at 1580 nm corresponds to the loss measured at 1554 nm in the experiment, where the loss is recorded as 1.54 dB. For the 1x2 and 1x3 metasplitter structures, shown in Figure 3-(g) and (h), the minimum measured loss values are -5.05 dB and -6.51 dB, respectively, which align well with the simulation results of -4.80 dB for the 1x2 metasplitter and -6.09 dB for the 1x3 metasplitter. The additional loss introduced by fabrication imperfections is minor in all cases compared to the simulation results at 1580 nm, the point of minimal loss in the metasplitter designs. It is also important to note that the observed band shifts in both the 1DGW and the output ports of the metasplitters are attributed to the sensitivity of the 1DGW structure to dimensional variations \cite{saghaei2022sinusoidal}, as well as the extremely small 100 nm feature size of the metastructures (Sect. S4, Supplement I).

As shown in the SEM images in Figure 3-(a) and (b), the effects of the 1DGW structure in terms of loss and bandwidth are also included in the transmission results provided in Figure 3-(d), (e), (g), and (h). However, to accurately assess the intrinsic performance of the metastructures, these 1DGW effects must be excluded from the transmission data. The resulting intrinsic transmission results for the metasplitters are presented in Figure 3-(i) and (j) for the 1x2 configuration, and in Figure 3-(m) and (n) for the 1x3 configuration. In Figure 3-(i), the simulated loss for the 1x2 metasplitter at 1580 nm is 4.08 dB, with a 1 dB bandwidth of 28 nm. In Figure 3-(j), the measured results for the same structure show a minimum loss of 3.80 dB and a 1 dB bandwidth of 15 nm, where the 1 dB bandwidth limits of both channels intersect. Similarly, Figure 3-(m) presents, for the 1x3 metasplitter, an average simulated minimum loss of 5.61 dB at 1580 nm with a 1 dB bandwidth of 29 nm, while Figure 3-(n) shows the measured average minimum loss of 5.36 dB with a 15 nm 1 dB bandwidth, where the 1 dB bandwidth regions of all channels intersect. These results indicate losses of only 0.80 dB and 0.59 dB compared to ideal splitting operations (-3 dB for 1x2 (Figure 3-(j)) and -4.77 dB for 1x3 (Figure 3-(n))). Notably, the measured intrinsic transmission values of the metasplitters are slightly higher than the simulated results. This occurs because, when the 1DGW effects are excluded, band shift differences in the measured data cause slight alterations in transmission, leading to the observed increase in the measured transmission. To observe the difference between channels, the uniformity is calculated using the formula \( U = -10 \cdot \log\left[\frac{\min(P_{1 \times N})}{\max(P_{1 \times N})}\right] \). As shown in Figure 3-(k) and (l) for the 1x2 metasplitter, the simulated uniformity is zero due to symmetrical conditions, while in the measurement results, it remains well below 1 dB across a wide bandwidth.  For the 1x3 metasplitter, as shown in Figure 3-(o) and (p), uniformity below 1 dB is consistently maintained across a wide bandwidth in both the simulation and measurement results.

In this Letter, we introduce the first Bloch mode metasplitter in 1x2 and 1x3 configurations for 1DGWs, designed to achieve the smallest possible footprint and low-loss wideband operation. The proposed study represents a novel structure for the photonic library and introduces genetic algorithm-based framework for increasingly prevalent time-domain-oriented inverse design approaches. While the proposed configurations are designed with the smallest possible footprint, the methodology is extendable to larger 1xN systems with diverse power distribution ratios. This study paves the way for advancements in complex optical antenna arrays and dense, high-bandwidth true-time-delay systems, fostering greater versatility and broader utilization of 1DGWs.

\section*{Acknowledgement}
We would like to thank Prof. Lukas Chrostowski for his insightful comments and discussion. We also express our gratitude to the Silicon Electronic-Photonic Integrated Circuits (SiEPIC) program for their support.

\section*{Funding}
This research was supported by the Scientific and Technological Research Council of Turkey (TUBITAK), project number 122E566. Ahmet O. Sakin acknowledges the support of the TUBITAK BIDEB 2210A grant.

\section*{Disclosures}
The authors declare no conflicts of interest.

\section*{Data Availability Statement} Data underlying the results presented in this Letter are not publicly available but may be obtained from the authors upon reasonable request.

\bibliographystyle{unsrt}  
\bibliography{references}  

\includepdf[pages=-]{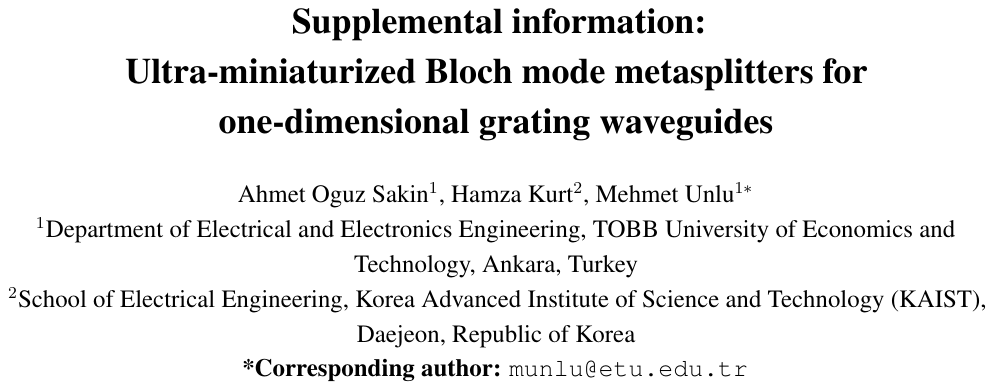}

\end{document}